%% file: mMIMO-U_ICC_camera_ready.tex
\documentclass[final]{IEEEtran}

\newlength \figwidth
\usepackage{cite}
\ifCLASSINFOpdf
  \usepackage[pdftex]{graphicx}
	\usepackage{epstopdf}
  \graphicspath{{../Figures/}}
  \DeclareGraphicsExtensions{.eps}
\else
  \usepackage[dvips]{graphicx}
\fi
\usepackage[cmex10]{amsmath}
\usepackage[tight,footnotesize]{subfigure}
\usepackage{amssymb}
\usepackage{amsthm}
\usepackage{url}
\usepackage{dsfont}
\usepackage{tabulary}
\usepackage{multirow}
\usepackage{color}
\newcounter{MYtempeqncnt}

\usepackage{color}
\usepackage{times}
\usepackage{verbatim}
\usepackage{floatflt}
\usepackage{enumerate}
\usepackage{array}
\usepackage{multicol,afterpage,wrapfig} 
\usepackage{tikz}
\usepackage{algorithm}
\usepackage[noend]{algpseudocode}
\makeatletter
\def\BState{\State\hskip-\ALG@thistlm}
\makeatother
\usepackage{amsmath,amssymb,eucal}
\usepackage[utf8]{inputenc}
\usepackage{epsfig}
\usepackage{exscale}
\usepackage{latexsym}
\usepackage{verbatim}
\usepackage{amsfonts}
\usepackage{subfigure}
\usepackage{multirow}
\usepackage{cite}
\usepackage{balance}
\usepackage{color}
\usepackage{transparent}
\usepackage{import}
\usepackage{mathtools}
\usepackage{tikz}
\usetikzlibrary{automata,arrows,positioning,calc}
\usepackage{bbm}
\usepackage[printonlyused,withpage]{acronym}
\usepackage{tabu}
\usepackage{longtable}
\usepackage{balance}
\usepackage{footnote}
\usepackage{tablefootnote}

\def\BibTeX{{\rm B\kern-.05em{\sc i\kern-.025em b}\kern-.08em
    T\kern-.1667em\lower.7ex\hbox{E}\kern-.125emX}}
    
\renewcommand{\IEEEQED}{\IEEEQEDopen}

\newcommand*\xbar[1]{%
  \hbox{%
    \vbox{%
      \hrule height 0.5pt 
      \kern0.36ex
      \hbox{%
        \kern-0.12em
        \ensuremath{#1}%
        \kern-0.12em
      }%
    }%
  }%
}

\input{macros}

\usetikzlibrary{arrows,calc}
\allowdisplaybreaks

\IEEEoverridecommandlockouts
\begin{document}
\pagenumbering{gobble}

\newtheorem{Theorem}{\bf Theorem}
\newtheorem{Corollary}{\bf Corollary}
\newtheorem{Remark}{\bf Remark}
\newtheorem{Lemma}{\bf Lemma}
\newtheorem{Proposition}{\bf Proposition}
\newtheorem{Assumption}{\bf Assumption}
\newtheorem{Definition}{\bf Definition}
\title{Enhancing Coexistence in the Unlicensed\\Band with Massive MIMO}
\author{\IEEEauthorblockN{{Giovanni~Geraci, Adrian~Garcia-Rodriguez, David~L\'{o}pez-P\'{e}rez, Andrea~Bonfante,\\Lorenzo~Galati~Giordano, and Holger Claussen}}\\
\normalsize\IEEEauthorblockA{\emph{Bell Laboratories Nokia, Ireland}}}
\maketitle
\thispagestyle{empty}
\IEEEpeerreviewmaketitle
\input{Abstract}
\input{Section1}
\input{Section2}
\input{Section3}
\input{Section4}
\input{Section5}
\ifCLASSOPTIONcaptionsoff
  \newpage
\fi
\bibliographystyle{IEEEtran}
\bibliography{Strings_Gio,Bib_Gio}
\end{document}

%% file: macros.tex
\newfont{\bbb}{msbm10 scaled 500}

\newfont{\bb}{msbm10 scaled 1100}

\newcommand{\executeiffilenewer}[3]{%
\ifnum\pdfstrcmp{\pdffilemoddate{#1}}%
{\pdffilemoddate{#2}}>0%
{\immediate\write18{#3}}\fi%
}

%% file: Abstract.tex
\begin{abstract}
We consider cellular base stations (BSs) equipped with a large number of antennas and operating in the unlicensed band. We denote such system as massive MIMO unlicensed (mMIMO-U). We design the key procedures required to guarantee coexistence between a cellular BS and nearby Wi-Fi devices. These include: neighboring Wi-Fi channel covariance estimation, allocation of spatial degrees of freedom for interference suppression, and enhanced channel sensing and data transmission phases. We evaluate the performance of the so-designed mMIMO-U, showing that it allows simultaneous cellular and Wi-Fi transmissions by keeping their mutual interference below the regulatory threshold. The same is not true for conventional listen-before-talk (LBT) operations. As a result, mMIMO-U boosts the aggregate cellular-plus-Wi-Fi data rate in the unlicensed band with respect to conventional LBT, exhibiting increasing gains as the number of BS antennas grows.
\end{abstract}

%

%% file: Section1.tex
\section{Introduction}
In view of the ever increasing mobile data demand, the wireless industry has turned its attention to unlicensed spectrum bands, e.g., 2.4 and 5 GHz, to provide extra resources for cellular networks \cite{Huawei2013,QualcommMulteFire2015,NokiaMulteFire2015,ZhaWanCai2015}. Although unlicensed spectrum allows mobile operators to serve more users via traffic offloading and/or to enhance their quality of service through carrier aggregation, harmonious coexistence with other technologies working in the unlicensed spectrum, such as IEEE~802.11x (Wi-Fi), must be guaranteed \cite{ZhaChuGuo2015,MukCheFal2015,BenSimCzy2013}. Wi-Fi systems rely on a contention-based access with a random backoff mechanism. Therefore, cellular base stations (BSs) transmitting continuously over the unlicensed bands would result in interference and repeated backoffs at the Wi-Fi nodes \cite{PerSta2013}.

Coexistence between cellular BSs and Wi-Fi devices is currently achieved with two possible approaches: carrier-sensing adaptive transmission (CSAT) and listen before talk (LBT). With CSAT, cellular BSs interleave their transmissions with idle intervals that allow Wi-Fi devices to access the channel \cite{LTEUForum:15,RahBehKoo2011}. With LBT, cellular BSs sense the channel via energy detection, and they commence a transmission in the unlicensed band only if the channel is deemed free for a designated time. In some regions, e.g., Europe and Japan, such channel sensing operation is mandatory \cite{3GPP-RP-140808}. While ensuring coexistence, neither CSAT nor LBT allow simultaneous usage of the unlicensed spectrum by both cellular BSs and Wi-Fi devices when the latter fall in the coverage area of the former.

In this paper, we propose massive multiple-input multiple-output (MIMO) as a means to enhance coexistence while maximizing spectrum reuse in the unlicensed band. Massive MIMO has recently emerged as one of the potential disruptive technologies for the fifth generation (5G) wireless systems, where cellular BSs are envisioned to be equipped with a large number of antennas \cite{Mar:10,RusPerBuoLar:2013,BjoLarDeb2016}. In particular, we consider a massive MIMO system operating in the unlicensed band. We refer to this system as massive MIMO unlicensed (mMIMO-U), where the spatial degrees of freedom (d.o.f.) provided by the large number of antennas are employed to suppress the interference generated towards Wi-Fi devices operating in the neighborhood of each massive MIMO BS. This allows massive MIMO BSs and Wi-Fi devices to use unlicensed bands simultaneously, thus increasing the network spatial reuse.

Our contributions can be summarized as follows.
\begin{itemize}
\item \textit{Design:} We devise the key operations of a mMIMO-U system, which include (i) neighboring Wi-Fi channel covariance estimation, (ii) spatial d.o.f. allocation and user scheduling, (iii) precoder calculation, and (iv) enhanced listen before talk and data transmission phases.
\item \textit{System-level simulations:} We evaluate the performance of the proposed mMIMO-U system in scenarios of practical interest, showing that it significantly reduces mutual interference between cellular BSs and Wi-Fi devices, and that it boosts the aggregate data rate per sector.
\item \textit{Discussion and insights:} We identify the main follow-ups of the present work, providing an overview of important issues related to cross-layer system modeling, channel estimation and scheduling procedures, as well as network deployment strategies in the unlicensed band.
\end{itemize}
 

%% file: Section2.tex
\section{System Set-Up}

We now provide a general introduction to the network topology and propagation model. More details on the parameters used for our system-level study will be given in Section~IV.

\begin{figure*}[!t]
\centering
\includegraphics[width=\figwidth]{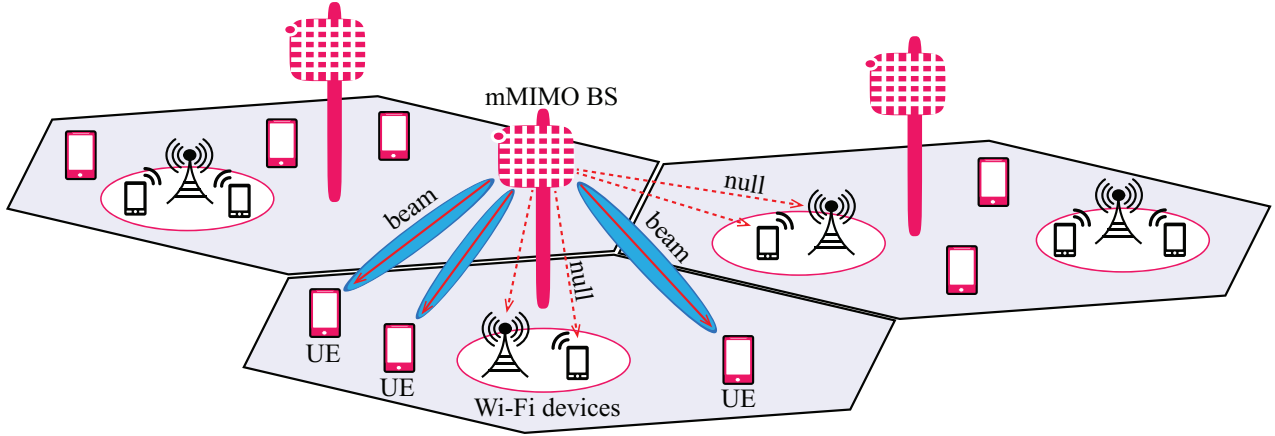}
\caption{Illustration of a mMIMO-U system: Each BS multiplexes UEs in the unlicensed band while suppressing interference at neighboring Wi-Fi devices.}
\label{fig:SystemModel}
\end{figure*}

\subsection{Network Topology}

We consider the downlink of a cellular network as shown in Fig.~\ref{fig:SystemModel}, where massive MIMO cellular BSs are deployed to operate in the unlicensed band and communicate with their respective sets of connected cellular user equipment (UEs). We assume that multiple Wi-Fi devices, i.e., access points (APs) and stations (STAs), are simultaneously operating in the same unlicensed band. As a result, BS-to-UE downlink transmissions may affect Wi-Fi transmissions, and vice versa.

We assume that cellular BSs and Wi-Fi devices transmit with power $P_{\textrm{b}}$ and $P_{\textrm{w}}$, respectively, and that cellular UEs (resp. Wi-Fi STAs) associate to the cellular BS (resp. Wi-Fi AP) that provides the largest average received power. Each BS $i$ is equipped with a large number of antennas $N$, and it simultaneously serves $K_i$ of its associated UEs, $K_i \leq N$, on each time-frequency resource block (RB) through spatial multiplexing. While the total number of associated UEs is determined by the UE density and distribution and by the nature of traffic, the value of $K_i$ can be chosen adaptively by BS $i$ via scheduling operations.

\subsection{Channel Model}

All propagation channels are affected by antenna gain, path loss, shadowing, and fast fading, as detailed in Section~IV, and are assumed constant within a time/frequency coherence block. Without loss of generality, we assume that all UEs and Wi-Fi devices are equipped with a single antenna.\footnote{In Section~III, the proposed Wi-Fi channel covariance estimation phase is general and holds irrespective of the number of Wi-Fi antennas, whereas the proposed scheduling and precoding operations can be modified to account for multi-antenna UEs served through multiple data streams.}
Throughout the paper, we employ $[m]$ as the discrete time variable, $(\cdot)^{\dagger}$ denotes conjugate transpose, and we define the following variables:
\begin{itemize}
\item $\mathbf{h}_{ijk} \in \mathbb{C}^{N\times 1}$ denotes the channel vector between BS $i$ and UE $k$ in cell $j$;
\item $\mathbf{g}_{i\ell} \in \mathbb{C}^{N\times 1}$ denotes the channel vector between BS $i$ and Wi-Fi device $\ell$;
\item $q_{\ell jk} \in \mathbb{C}$ denotes the channel coefficient between Wi-Fi device $\ell$ and UE $k$ in cell $j$.
\end{itemize}

The signal $y_{ik}[m] \in \mathbb{C}$ received by UE $k$ in cell $i$ is given by (\ref{eqn:rx_signal}) at the top of next page, where (i) $\mathbf{w}_{ik} \in \mathbb{C}^{N\times 1}$ is the precoding vector from BS $i$ to UE $k$ in cell $i$, (ii) $s_{ik}[m] \in \mathbb{C}$ is the signal intended for UE $k$ in cell $i$, (iii) $s_{\ell}[m] \in \mathbb{C}$ is the signal transmitted by Wi-Fi device $\ell$, and (iv) $\epsilon_{ik}[m]$ is the thermal noise. The four above quantities respectively satisfy the following: (i) $\sum_{k=1}^{K_i} \Vert \mathbf{w}_{ik}\Vert^2 = 1 \enspace\forall i$, (ii) $\mathbb{E}\left[ \vert s_{ik}[m] \vert^2\right] = 1$, (iii) $\mathbb{E}\left[ \vert s_{\ell}[m] \vert^2\right] = 1$, and (iv) $\epsilon_{ik}[m] \sim \mathcal{CN}(0,\sigma^2_{\epsilon})$. We note that the five terms on the right hand side of (\ref{eqn:rx_signal}) respectively represent: useful signal, intra-cell interference from the serving BS, inter-cell interference from other BSs, interference from Wi-Fi devices, and thermal noise. The resulting SINR $\nu_{ik}$ at UE $k$ in cell $i$ is given by (\ref{equ:sinr_ue}) at the top of next page.

\begin{figure*}[!t]
\normalsize
\setcounter{MYtempeqncnt}{\value{equation}}
\hrulefill
\begin{align}
y_{ik}[m] = \!\sqrt{P_{\textrm{b}}} {\mathbf{h}_{iik}^{\dagger} \mathbf{w}_{ik} s_{ik}[m]} + \!\!\!\!\!\!\! \sum_{k'=1, k' \neq k}^{K_i} \!\!\!\!\!\!\! \sqrt{P_{\textrm{b}}} {\mathbf{h}_{iik}^{\dagger} \mathbf{w}_{ik'} s_{ik'}[m]}
+ \sum_{i' \neq i} \sum_{k'=1}^{K_{i'}} \!\! \sqrt{P_{\textrm{b}}} {\mathbf{h}_{i'ik}^{\dagger} \mathbf{w}_{i'k'} s_{i'k'}[m]} \!+\! \sum_{\ell} \! \sqrt{P_{\textrm{w}}} {q_{\ell ik} s_{\ell}[m]} \!+\! \epsilon_{ik}[m]
\label{eqn:rx_signal}
\end{align}
\setcounter{equation}{\value{MYtempeqncnt}}
\addtocounter{equation}{1}
\setcounter{MYtempeqncnt}{\value{equation}}
\begin{align}\label{equ:sinr_ue}
\nu_{ik} = \frac{ P_{\textrm{b}} \vert \mathbf{h}_{iik}^{\dagger} \mathbf{w}_{ik} \vert^2 }
{P_{\textrm{b}} \sum_{k'=1,k' \neq k}^{K_i} \vert \mathbf{h}_{iik}^{\dagger} \mathbf{w}_{ik'} \vert^2 + 
P_{\textrm{b}} \sum_{i' \neq i} \sum_{k'=1}^{K_{i'}} \vert \mathbf{h}_{i'ik}^{\dagger} \mathbf{w}_{i'k'} \vert^2 + 
P_{\textrm{w}} \sum_{\ell} \vert q_{\ell ik} \vert^2
+ \sigma^2_{\epsilon}}
\end{align}
\hrulefill
\setcounter{equation}{\value{MYtempeqncnt}}
\vspace*{-3pt}
\end{figure*}
\addtocounter{equation}{1}

\subsection{Coexistence with Wi-Fi Devices}

When mMIMO BSs operate in the unlicensed band, all Wi-Fi devices receive interfering signals. The interference power $I_{\ell}[m]$ received at Wi-Fi device $\ell$ due to mMIMO-U downlink operations is given by
\begin{equation}
I_{\ell}[m] = P_{\textrm{b}} \sum_{i} \sum_{k=1}^{K_{i}} \left| \mathbf{g}_{il}^{\dagger} \mathbf{w}_{ik} s_{ik}[m] \right|^2.
\label{eqn:Il}
\end{equation}

Each Wi-Fi device $\ell$ deems the channel as occupied and defers from transmission when the total received power, i.e., from all mMIMO-U BSs and all Wi-Fi devices, falls above a certain threshold. As result, controlling the interference power in (\ref{eqn:Il}) generated by the mMIMO-U BSs and keeping it to a minimum would facilitate seamless coexistence, i.e., Wi-Fi device $\ell$ would continue using the wireless medium as if no mMIMO-U transmissions were taking place.

%% file: Section3.tex
\section{Massive MIMO Unlicensed Operations}

In current coexistence approaches, each BS performs an LBT phase before any data transmission, which consists of measuring the total power received from Wi-Fi devices. A transmission opportunity is then gained by the BS if the sum power received from all devices using the same band falls below a certain threshold for a designated time interval. This process is also known as energy detection. Note that this allows for either a single BS or a Wi-Fi device to transmit within a certain coverage area, thus preventing spatial reuse of the same unlicensed band. The above is true irrespective of the number of antennas at the cellular BS.

In the proposed mMIMO-U system, each BS exploits the large number of transmit antennas to enhance coexistence with neighboring Wi-Fi devices operating in the unlicensed band, so that both cellular BSs and Wi-Fi devices can simultaneously use the unlicensed band. The main operations we propose to perform at the mMIMO-U BS can be outlined as follows: (i) UE and Wi-Fi channel estimation, (ii) allocation of spatial d.o.f. for interference suppression and scheduling, (iii) precoder calculation, and (iv) enhanced LBT and data transmission. We now provide a detailed description for each of the above four operations.

\subsection{Channel Estimation}

In order to spatially multiplex the served UEs, each BS requires knowledge of their channels $\mathbf{h}_{iik}$. UE channel state information (CSI) may be obtained via pilot signals transmitted during a training phase. Pilot signals may be interfered with by concurrent Wi-Fi transmissions due to hidden-terminal issues, and may suffer pilot contamination due to pilot reuse across mMIMO-U cells. Interference received during the training phase yields an imperfect UE CSI estimation, which may reduce the ability of each BS to spatially multiplex the served UEs, thus degrading their SINR. We denote as $\widehat{\mathbf{h}}_{iik}$ the imperfect CSI available at BS $i$ for UE $k$ in cell $i$.

In order to suppress interference at the neighboring Wi-Fi devices, each BS $i$ requires information on their channels $\mathbf{g}_{i\ell}$. In our proposed mMIMO-U system, when BS $i$ is not transmitting data, it obtains the channel subspace spanned by $\mathbf{g}_{i\ell}$ through a covariance estimation procedure. In fact, when silent, BS $i$ receives the following signal
\begin{equation}
\mathbf{z}_i[m] = \sqrt{P_{\mathrm{w}}} \sum_{\ell} \mathbf{g}_{i\ell} s_{\ell}[m] + \boldsymbol{\eta}_i[m],
\label{eqn:zi}
\end{equation}
which consists of all Wi-Fi transmissions and of a noise term $\boldsymbol{\eta}_i[m]$. We note that $\boldsymbol{\eta}_i[m]$ may include not only thermal noise, but also signals received from other BSs if they choose their silent periods in an asynchronous, non-overlapped, fashion.\footnote{In the case of other cellular operators using the same unlicensed band, (\ref{eqn:zi}) would include their transmitted signals, and the mMIMO-U system would ensure coexistence with these operators as well as with Wi-Fi devices.} We neglect the presence of such signals, and assume $\boldsymbol{\eta}_i[m] \sim \mathcal{CN}(\mathbf{0},\sigma^2_{\eta} \mathbf{I})$. Let $\mathbf{Z}_i = \mathbb{E} [ \mathbf{z}_i \mathbf{z}_i^{\dagger}] \in \mathbb{C}^{N \times N}$ be the covariance of $\mathbf{z}_i[m]$, where the expectation is taken with respect to $s$ and $\boldsymbol{\eta}$. BS $i$ can obtain an estimate $\widehat{\mathbf{Z}}_i$ of $\mathbf{Z}_i$ via a simple average over $M$ discrete time intervals as \cite{HoyHosTen2014}
\begin{equation}
\widehat{\mathbf{Z}}_i = \frac{1}{M} \sum_{m=1}^M \mathbf{z}_i[m] \mathbf{z}_i^{\dagger}[m].
\label{eqn:Z_hat}
\end{equation}

\subsection{Spatial Resource Allocation and Scheduling}

From the Wi-Fi covariance estimate, BS $i$ calculates the number of spatial d.o.f. $D_{i}$ that will be allocated to suppress interference at the neighboring Wi-Fi devices. We note that a sufficient number of d.o.f. must be allocated for interference suppression, such that the interference generated by the BS at the Wi-Fi devices does not exceed a certain threshold, thus ensuring coexistence.

Given the covariance matrix $\widehat{\mathbf{Z}}_i$, BS $i$ applies a spectral decomposition, obtaining
\begin{equation}
\widehat{\mathbf{Z}}_i = \widehat{\mathbf{U}}_i \widehat{\mathbf{\Lambda}}_i \widehat{\mathbf{U}}_i^{\dagger},
\label{eqn:spectral}
\end{equation}
where the columns of $\widehat{\mathbf{U}}_i=[\widehat{\mathbf{u}}_{i1},\ldots,\widehat{\mathbf{u}}_{iN}]$ form an orthonormal basis and
\begin{equation}
\widehat{\mathbf{\Lambda}}_i = \textrm{diag}\left( \widehat{\lambda}_{i1},\ldots,\widehat{\lambda}_{iN} \right)	
\end{equation}
contains a set of eigenvalues, such that $\widehat{\lambda}_{i1} \geq \widehat{\lambda}_{i2} \ldots \geq \widehat{\lambda}_{iN}$. BS $i$ then suppresses interference on the subspace spanned by the $D_{i}$ dominant eigenvectors $[\widehat{\mathbf{u}}_{i1},\ldots,\widehat{\mathbf{u}}_{iD_i}]$, where a variety of criteria can be employed to choose the value of $D_{i}$. We provide two illustrative examples in the following.

\subsubsection{Fixed number of scheduled UEs}
A scheduler chooses the number of UEs $K_i\leq N$ to be served in the unlicensed band. The value of $D_{i}$ cannot exceed the remaining spatial d.o.f. at the BS, and it is obtained as
\begin{equation}
D_{i} = \lfloor c_1 \left( N - K_i \right) \rfloor,
\label{eqn:Di_fixed_Ki}
\end{equation}
where $\lfloor \cdot \rfloor$ denotes the floor function, and $0 < c_1 < 1$ controls the fraction of excess d.o.f. used for interference suppression, thus trading off precoding gain at the UEs for enhanced interference suppression at the Wi-Fi devices. The results shown in Section~IV are obtained by using the above procedure.

\subsubsection{Controlled interference at Wi-Fi devices}
The value of $D_{i}$ is chosen as the number of eigenvalues that satisfy $\widehat{\lambda}_{in} > \gamma$, so that the following holds for the aggregate interference generated by BS $i$ at all Wi-Fi devices
\begin{equation}
\begin{aligned}
\left\| \sum_{k=1}^{K_i} \sqrt{P_{\mathrm{b}}} \mathbf{U}_i^{\dagger} \mathbf{\Lambda}_i^{\frac{1}{2}} \mathbf{w}_{ik} \right\|^2 &\overset{\textrm{(a)}}{\approx} \left\| \sum_{k=1}^{K_i} \sqrt{P_{\mathrm{b}}} \widehat{\mathbf{U}}_i^{\dagger} \widehat{\mathbf{\Lambda}}_i^{\frac{1}{2}} \mathbf{w}_{ik} \right\|^2\\
&\overset{\textrm{(b)}}{\leq} P_{\mathrm{b}} \widehat{\lambda}_{i,{D_{i}+1}} \leq P_{\mathrm{b}} \gamma,
\end{aligned}
\label{eqn:interference}
\end{equation}
where (a) follows from (\ref{eqn:Z_hat}), and (b) holds by considering the worst case where all precoders are aligned with $\widehat{\mathbf{u}}_{i,D_i+1}$. The threshold $\gamma \geq 0$ trades off the amount of interference generated at the Wi-Fi devices with the number of d.o.f. spent for interference suppression. A scheduler then chooses the number of UEs $K_i$ as
\begin{equation}
K_i = \lfloor c_2 \left( N - D_{i} \right) \rfloor,
\end{equation}
where $0 < c_2 < 1$ controls the fraction of excess d.o.f. used to multiplex UEs, thus trading off precoding gain and multiplexing gain at the UEs.\footnote{See, e.g., \cite{BjoLarDeb2016}, for relevant discussions on the choice of $K_i$ versus $N$.}

\subsection{Precoder Calculation}

Thanks to the plurality of transmit antennas, BS $i$ is able to spatially multiplex $K_i$ UEs while forcing $D_i$ nulls on the channel subspace spanned by the neighboring Wi-Fi devices, as depicted in Fig.~\ref{fig:SystemModel}. By defining $\mathbf{S}_i \in \mathbb{C}^{N\times (K_i+D_i)}$ as
\begin{equation}
\mathbf{S}_i = \left[ \widehat{\mathbf{h}}_{ii1},\ldots,\widehat{\mathbf{h}}_{ii{K_i}},\widehat{\mathbf{u}}_{i1},\ldots,\widehat{\mathbf{u}}_{iD_i} \right],
\label{eqn:Si}
\end{equation}
the precoding vector $\mathbf{w}_{ik}$ between BS $i$ and UE $k$ in cell $i$, $k=1,\ldots,K_i$, is obtained as follows
\begin{align}
\mathbf{w}_{ik} = \frac{1}{\sqrt{\zeta_i}} \mathbf{S}_i \left( \mathbf{S}_i^{\dagger} \mathbf{S}_i \right)^{-1} \mathbf{v}_k,
\label{eqn:precoder}
\end{align}
where $\mathbf{v}_k$ is a vector whose $k$-th entry is unitary and the remaining entries are zero, and where the constant $\zeta_i$  is chosen to normalize the average transmit power such that
\begin{equation}
\sum_{k=1}^{K_i}\Vert \mathbf{w}_{ik}\Vert^2 = 1.	
\end{equation}

Note that in (\ref{eqn:Si}), the vectors $[\widehat{\mathbf{u}}_{i1},\ldots,\widehat{\mathbf{u}}_{iD_i}]$ span the channel subspace on which BS $i$ receives a significant Wi-Fi-transmitted power. 
Wi-Fi uplink/downlink and BS downlink transmissions share the same frequency band, hence, due to channel reciprocity, any power transmitted by BS $i$ on such subspace would generate significant interference at one or more Wi-Fi devices. For this reason, BS $i$ sacrifices $D_i$ spatial d.o.f. to suppress the interference generated on the directions $[\widehat{\mathbf{u}}_{i1},\ldots,\widehat{\mathbf{u}}_{iD_i}]$. Such interference suppression is performed by all BSs in a distributed manner, and it improves coexistence with the Wi-Fi devices operating in the unlicensed band, as it will be shown in Section~IV.

\subsection{Enhanced LBT and Data Transmission}

In order to comply with the regulations in the unlicensed band, each BS must perform LBT before any data transmission \cite{3GPP-RP-140808}. With conventional LBT operations, a transmission opportunity is gained by the BS if the total measured power in (\ref{eqn:zi}) satisfies $\| \mathbf{z}_i[m]\|^2 < \gamma_{\mathrm{LBT}}$ for a designated time interval. In the proposed mMIMO-U system, the LBT phase is enhanced as follows. When BS $i$ listens to the transmissions currently taking place in the unlicensed band, it filters the received signal $\mathbf{z}_i[m]$ with the $D_i$ spatial nulls in place. A transmission opportunity is then gained by the BS if the condition
\begin{equation}
\sum_{n=D_i+1}^{N} \left| \widehat{\mathbf{u}}_{in}^{\dagger} \mathbf{z}_i[m] \right|^2 < \gamma_{\mathrm{LBT}}
\label{eqn:LBT}
\end{equation}
holds for a designated time interval.

In other words, since BS $i$ aims to transmit downlink signals on a specific channel subspace, i.e., the one spanned by $[\widehat{\mathbf{u}}_{i,D_i+1},\ldots,\widehat{\mathbf{u}}_{iN}]$ and orthogonal to $[\widehat{\mathbf{u}}_{i1},\ldots,\widehat{\mathbf{u}}_{iD_i}]$, it must make sure that no concurrent transmissions are detected on that subspace. This is accomplished by measuring the aggregate power of the received signal filtered through the vectors $\mathbf{u}_{in}$, $n=D_i+1,\ldots,N$. Provided that a sufficient number of d.o.f. $D_i$ have been allocated for interference suppression, the condition in (\ref{eqn:LBT}) is met. Therefore, unlike conventional LBT operations, the enhanced LBT phase allows both mMIMO-U BSs and Wi-Fi devices to simultaneously use the unlicensed band.

%% file: Section4.tex
\section{System-Level Simulations}

In this section, we compare the mMIMO-U operations proposed in Section~III to a conventional LBT approach, where no Wi-Fi covariance estimation, enhanced LBT, nor interference suppression are performed. The behavior of both schemes is evaluated via system-level simulations, with an identical number of cellular BS antennas, and according to the scenario and methodologies described in Table~\ref{table:parameters}.

\begin{table}
\centering
\caption{System-Level Simulation Parameters}
\label{table:parameters}
\begin{tabulary}{\columnwidth}{ |l | L | }
\hline
    \textbf{Parameter} 			& \textbf{Description} \\ \hline
    Cellular layout				& Hexagonal with wrap-around, 19 sites, 3 sectors each, 1 cellular BS per sector \\ \hline
    Inter-site distance 		& 500m \cite{3GPP36814} \\ \hline
  UEs per sector 					& Random (P.P.P.), $K_i=8$ on average \\ \hline
	UEs distribution	& Distance from Wi-Fi hotspots: $d\geq 60~\text{m}$, distance from BSs: $25~\text{m}\leq d\leq 150\text{m}$\ifx\[$\else\tablefootnote{A scheduler can be employed at the BS to enforce this condition. UEs that do not meet this requirement can be served in the licensed band \cite{GerGarLop2016}.}\fi \\ \hline
	Wi-Fi hotspots \phantom{$\Big($}  & 2 outdoor hotspots per sector, radius: 20~m\ifx\[$\else\tablefootnote{We consider outdoor Wi-Fi devices since this case involves no wall penetration losses, making coexistence with cellular BSs more challenging.}\fi\\ \hline
	Wi-Fi devices  		& 8 devices per hotspot: 1~AP and 7~STAs \\ \hline
	Carrier frequency 		& 5.15 GHz (U-NII-1) \cite{FCC1430} \\ \hline
	System bandwidth 			& 20 MHz with 100 resource blocks \cite{FCC1430, 3GPP36814} \\ \hline
	Wi-Fi throughput  	& 65 Mbps per cluster \cite{PerSta2013}\\ \hline
	LBT regulations  		& Threshold $\gamma_{\mathrm{LBT}}=-62$ dBm \cite{PerSta2013} \\ \hline    
	d.o.f. allocation		& $D_i$ as per (\ref{eqn:Di_fixed_Ki}) with $c_1=0.5$ \\ \hline    
	BS precoder 				& mMIMO-U: as in (\ref{eqn:precoder}), LBT: zero forcing \cite{Wagner12} \\ \hline
	BS antennas			& downtilt: $12^{\circ}$, height: $25$~m  \cite{3GPP36814} \\ \hline
	BS antenna array 			& Uniform linear, element spacing: $d = 0.5\lambda$ \\ \hline
	BS antenna pattern 			& Urban macro (6 dBi max.) \cite{3GPP36814} \\ \hline
	BS tx power 		& 30 dBm \cite{FCC1430} \\ \hline   
	Wi-Fi tx power 		& APs: 24~dBm, STAs: 18~dBm \cite{FCC2013}, average per cluster: 19.36~dBm \\ \hline   
	UE noise figure 			& 9 dB \cite{3GPP36824} \\ \hline
	UE rx sensitivity 			& -94 dBm \cite{3GPP36101} \\ \hline
    Fast fading  				& Ricean, distance-dependent K factor \cite{4804716} \\ \hline
	Lognormal shadowing 					& BS to UE as per \cite{3GPP36814}, UE to UE as per \cite{3GPP36843} \\ \hline
	Channel correlation	& Jakes correlation model \cite{656151} \\ \hline
	Path loss 					& 3GPP UMa \cite{3GPP36814} and 3GPP D2D \cite{3GPP36843}  \\ \hline
	CSI error 	& $\mathbf{\widehat{h}}_{iik} = \sqrt{1 - \tau^2} \mathbf{h}_{iik} + \tau \mathbf{e}_{iik}$, where $\mathbf{e}_{iik} \sim 									  \mathcal{CN} \left( \mathbf{0}, \mathbf{I}_N \right)$ and $\tau^2 = 0.1$ \cite{Wagner12}. \\ \hline
	Thermal noise & -174 dBm/Hz spectral density\\ \hline
\end{tabulary}
\end{table}

\subsection{Enhanced Coexistence}

\begin{figure}[!t]
\centering
\includegraphics[width=\columnwidth]{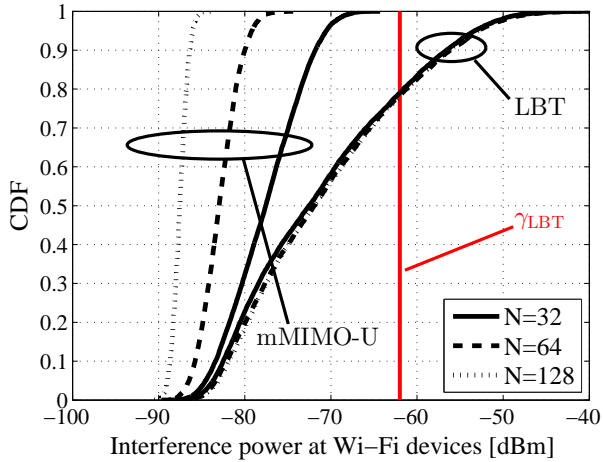}
\caption{Coexistence in the unlicensed band as seen by Wi-Fi devices.}
\label{fig:WiFiInt}
\end{figure}

Figure~\ref{fig:WiFiInt} shows coexistence in the unlicensed band from the perspective of Wi-Fi devices (both APs and STAs), comparing mMIMO-U to conventional LBT. It is assumed that cellular BSs have gained access to the unlicensed medium, and the cumulative distribution function (CDF) of the interference received by Wi-Fi devices is shown. It is evident from Fig.~\ref{fig:WiFiInt} that with mMIMO-U, Wi-Fi devices are able to use the unlicensed band while cellular BSs are transmitting, since the aggregate interference $I_{\ell}[m]$ is always below the regulatory threshold $\gamma_{\mathrm{LBT}} = -62$~dBm. On the other hand, with conventional LBT, Wi-Fi devices backoff 21\% of the time because the interference they receive is above $\gamma_{\mathrm{LBT}}$. Moreover, Fig.~\ref{fig:WiFiInt} shows that even when below the threshold, the interference received with conventional LBT is 50\% of the time above $-72$~dBm, which affects the quality of Wi-Fi transmissions \cite{jindal2015lte}. This phenomenon is not observed with mMIMO-U.

\begin{figure}[!t]
\centering
\includegraphics[width=\columnwidth]{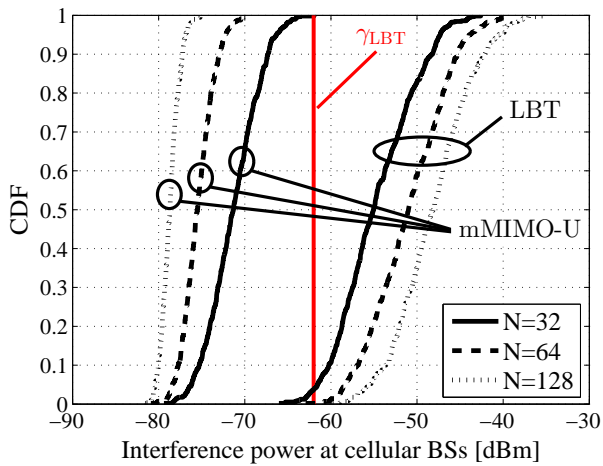}
\caption{Coexistence in the unlicensed band as seen by cellular BSs.}
\label{fig:BSInt}
\end{figure}

Figure~\ref{fig:BSInt} evaluates coexistence from the cellular BSs' standpoint, with mMIMO-U and with conventional LBT. It is assumed that Wi-Fi devices have gained access to the unlicensed medium, and the CDF of the interference received by cellular BSs is shown. Figure~\ref{fig:BSInt} confirms that cellular BSs implementing the proposed mMIMO-U are able to access the unlicensed band while Wi-Fi devices are transmitting. With $N = 32$ antennas, the aggregate interference received by cellular BSs is $100\%$ of the time below the threshold $\gamma_{\mathrm{LBT}}$. On the other hand, cellular BSs with conventional LBT incur repeated backoffs, since the interference they receive is $96\%$ of the time above $\gamma_{\mathrm{LBT}}$. Moreover, increasing the value of $N$ yields a larger interference at the cellular BSs with conventional LBT, because more aggregate power is received. Instead, the proposed mMIMO-U drastically reduces such interference for increasing $N$, since some of the excess d.o.f. are allocated for interference suppression as per (\ref{eqn:Di_fixed_Ki}). 

\subsection{Increased Data Rates}

\begin{figure}[!t]
\centering
\includegraphics[width=\columnwidth]{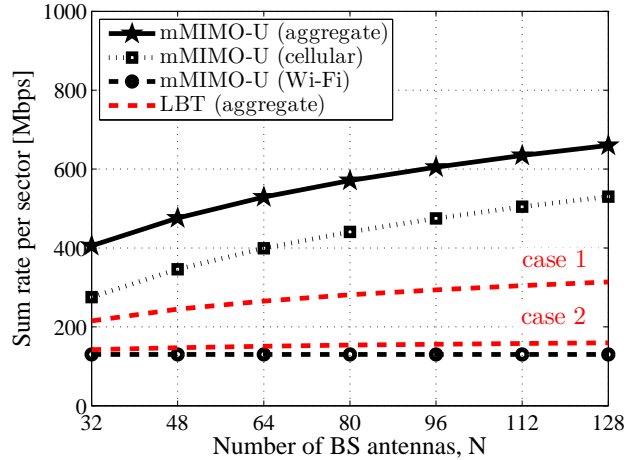}
\caption{Data rates with proposed mMIMO-U system versus conventional LBT. For LBT curves, case 1 and case 2 refer to Wi-Fi clusters operating in downlink only, and in both uplink and downlink, respectively.}
\label{fig:Rates}
\end{figure}

Figure~\ref{fig:Rates} shows the aggregate data rates per unit area with both mMIMO-U and conventional LBT. The rates of the mMIMO-U system account for the fact that cellular BSs and Wi-Fi devices do not transmit when their respective LBT phases are unsuccessful, and neglect the overhead associated with the covariance estimate in (\ref{eqn:Z_hat}). The rates plotted for conventional LBT refer to two cases: (i) Wi-Fi clusters operate in downlink only, such that a cellular BS shares the medium with the 2 Wi-Fi APs in its coverage sector; (ii) Wi-Fi clusters operate in both uplink and downlink, such that a BS shares the medium with all 16 Wi-Fi devices in its sector. Moreover, all rates provided by cellular BSs account for the receiver sensitivity, i.e., no data rate is achievable at a UE when the received power falls below the sensitivity threshold. Finally, Wi-Fi inter-cluster interference and collisions are neglected, and all rates provided by Wi-Fi APs are assumed equal to 65~Mbps when they gain access to the channel \cite{PerSta2013}.\footnote{As discussed in Section~V-B, a more accurate characterization of Wi-Fi rates in the presence of mMIMO-U transmissions requires higher-layer traffic models, e.g., those accounting for medium access control (MAC) protocols.}

The following observations can be made from Fig.~\ref{fig:Rates}. First, the Wi-Fi rates achieved by mMIMO-U are constant across all values of $N$ and equal to 130~Mbps per sector. This reflects the fact that devices from both Wi-Fi clusters in the sector can access the medium 100\% of the time, since the received interference is always below $\gamma_{\mathrm{LBT}}$ as shown in Fig.~\ref{fig:WiFiInt}.

Second, cellular rates with mMIMO-U are affected by the number of BS antennas $N$. For example, while 275~Mbps are achieved with $N=32$, cellular rates grow to 400~Mbps and 500~Mbps by increasing $N$ to 64 and 112, respectively. In fact, a larger number of antennas allows to suppress interference at Wi-Fi devices while leaving more spatial degrees of freedom to multiplex cellular UEs with a larger array gain. 

Third, mMIMO-U systems can boost the aggregate data rate in the unlicensed band. Indeed, although conventional LBT achieves up to 314~Mbps per sector, marginal gains are observed as the number of antennas $N$ is increased. This suggests that simply combining massive MIMO technology with conventional LBT operations is not performance effective. On the other hand, thanks to an enhanced coexistence between cellular BSs and Wi-Fi devices, mMIMO-U can guarantee more than double the aggregate rate, i.e., above 660~Mbps with $N=128$, also exhibiting increasing gains as $N$ grows.

%% file: Section5.tex
\section{Conclusion}

\subsection{Summary of Results}

We proposed a mMIMO-U system to enhance coexistence in the unlicensed band, and designed its main operations. These include: neighboring Wi-Fi channel covariance estimation, allocation of spatial d.o.f. for interference suppression, precoder calculation, and enhanced listen before talk. We considered a scenario of practical interest, and evaluated the performance of mMIMO-U through system-level simulations. Our results showed that mMIMO-U allows cellular BSs and Wi-Fi devices to use unlicensed bands simultaneously, thus increasing the network reuse compared to conventional LBT approaches.

\subsection{Discussion}

This work is suitable for several extensions from the system model, design, operations, and deployment perspectives:
\begin{itemize}
\item \textit{Model:} Accurate traffic models are desirable for Wi-Fi devices and multiple operators sharing the same unlicensed band, to evaluate how well BSs can estimate the channel covariance in (\ref{eqn:Z_hat}). The rate computation at Wi-Fi devices should also account for these traffic models, since even when the received interference falls below the threshold $\gamma_{\mathrm{LBT}}$, it may still affect the data rates \cite{jindal2015lte}.
\item \textit{Design:} Detailed procedures for UE CSI estimation should be defined, in order to guarantee coexistence between uplink pilots sent by UEs and Wi-Fi transmissions. One possible way to accomplish this would be to have BSs obtain access to the medium and reserve it for their UEs for the duration of a training phase only \cite{GerGarLop2016}.
\item \textit{Operations:} Given their CSI, scheduling operations are needed to select a group of $K_i$ UEs to be served. Conventional proportional fair (PF) schedulers may select UEs lying on the channel subspace spanned by the neighboring Wi-Fi devices. However, in mMIMO-U it may be desirable to select UEs whose channel is semi-orthogonal to such subspace, thus facilitating spatial multiplexing and interference suppression. UEs that do not satisfy this condition could be scheduled in the licensed band \cite{GerGarLop2016}.
\item \textit{Deployment:} Power emissions in the unlicensed band are strictly regulated. In some countries, the maximum allowed transmit power decreases with the number of antenna elements, if these are employed to focus energy in a particular direction \cite{FCC2013}. This means that the coverage area of mMIMO-U BSs may be limited. Therefore, an alternative strategy may consist of a more dense deployment of smaller low-power BSs, equipped with fewer antennas, covering smaller areas, and thus having to coexist with fewer Wi-Fi devices.
\end{itemize}